\documentclass[10pt]{article}
\usepackage[dvips]{graphicx}
  
  \newcommand{\lsim}{\raisebox{-4pt}{$\,\stackrel{\textstyle
							   <}{\sim}\,$}}
  \newcommand{\gsim}{\raisebox{-4pt}{$\,\stackrel{\textstyle
							   >}{\sim}\,$}}
  
  \newcommand{\be}{\begin{equation}}
  \newcommand{\ee}{\end{equation}}
  \newcommand{\ba}{\begin{eqnarray}}
  \newcommand{\ea}{\end{eqnarray}}
  \newcommand{\req}[1]{(\ref{#1})}
  \def\={\,=\,}
  \newcommand{\ci}[1]{\cite{#1}}

  \def\gev{~{\rm GeV}}

  \newcommand{\tw}{\textwidth}

  \def\vb0{{\bf b}_0}

  \def\={\,=\,}

\begin{document} 
  %%%%%%%%%%%%%%%%%%%%%%%%%%%%%%%%%%%%%%%%%%%%%%%%%%%%%%%%%%%%%%%%%%%%%%%%%%%%%%%%%%%%%%%%%%%%%
\thispagestyle{empty}
\begin{flushright}
WU B 09-11 \\
  %hep-ph/yymmnnn\\
October, 15 2009\\[20mm]
 \end{flushright}
%\markboth{Kroll}{Spin effects}

\vspace*{-0.05\tw}
 \begin{center}
{\Large\bf SPIN EFFECTS IN HARD EXCLUSIVE\\[0.2em] 
ELECTROPRODUCTION OF MESONS} 
  \vskip 8mm

  P. KROLL
  \\[1em]
  {\small \it Fachbereich Physik, Universit\"at Wuppertal, D-42097 Wuppertal,
  Germany\\
  and\\
  Institut f\"ur Theoretische Physik, Universit\"at
      Regensburg, \\D-93040 Regensburg, Germany\\
 Email:  kroll@physik.uni-wuppertal.de}
\end{center}
  %\maketitle
 
 \vskip 3mm 
  \begin{abstract}
  In this talk various spin effects in hard exclusive electroproduction of mesons
 are briefly reviewed. The data are discussed in the light of recent
 theoretical calculations within the frame work of the handbag approach.
 This talk has been presented at the Conference in Honor of Prof.\ Anatoly Efremov's
75th Birthday held at Trento, July, 2009. 
%\keywords{keyword1,2,3}  
\end{abstract}

\centerline{PACS Nos. 12.38.Bx, 13.60.Le, 13.88.+e}

%%%%%%%%%%%%%%%%%%%%%%%%%%%%%%%%%%%%%%%%%%%%%%%%%%%%%%%%%%%%%%%%%%%%%%%%%%%%%%%
  \section{Introduction}
  \label{sec:intro}
%%%%%%%%%%%%%%%%%%%%%%%%%%%%%%%%%%%%%%%%%%%%%%%%%%%%%%%%%%%%%%%%%%%%%%%%%%%
Electroproduction of mesons allows for the measurement of many spin effects. For 
instance, one may measure the dependence of the cross sections on the polarization
of the virtual photon by separation. Through the decay of vector mesons, e.g.\ 
$\rho^0\to\pi^+\pi^-$, one can measure the spin density matrix elements (SDME) of 
the decaying meson which also provide a wealth of information on spin effects. 
Last not least one may  work with longitudinally or transversely polarized targets 
and/or longitudinally polarized beams and measure various spin asymmetries. 
The investigation of spin-dependent observables allows for a deep insight in 
the underlying dynamics. Provided a sufficient number of them has been
measured the strength of the various contributing amplitudes and even their
relative phases can be determined from the experimental data. Here, in this 
article, it will be reported upon some spin effects and their dynamical 
interpretation in the frame work of the so-called handbag approach which 
offers a partonic description of meson electroproduction provided the 
virtuality of the exchanged photon, $Q^2$, is sufficiently large. The 
theoretical basis of the handbag approach is the factorization of the process 
amplitudes into a hard partonic subprocess and in soft hadronic matrix
elements, the so-called generalized parton distributions (GPDs) as well 
as wave functions for the produced mesons, see Fig.\ \ref{fig:0}. In collinear 
approximation factorization has been shown to hold rigorously for hard exclusive 
meson electroproduction.\ci{rad96,col96} It has been also shown by these authors 
that the transitions from a longitudinally polarized photon to a likewise 
polarized vector meson or a pseudoscalar one, $\gamma^*_L\to V_L (P)$, dominates 
for large $Q^2$. Other photon-meson transitions are suppressed by inverse
powers of the hard scale.

As mentioned spin effects in hard exclusive meson electroproduction will be 
briefly reviewed and their implications on the handbag approach and above all for the
determination of the GPDs, discussed. In Sect.\ 2 evidences for contributions from 
transversely polarized photons in vector-meson production are introduced. Next, in 
Sect.\ 3, the role of target spin asymmetries in meson electroproduction is examined 
and results for vector mesons shown. In Sect.\ 4 an estimated of the
GPD $E$, needed for a calculation of the target spin asymmetries for
vector mesons, is presented. Sect.\ 5 is devoted to a discussion of the the 
target spin asymmetries in pion electroproduction. Finally, in Sect.\ 6, a summary
is given. 

%%%%%%%%%%%%%%%%%%%%%%%%%%%%%%%%%%%%%%%%%%%%%%%%%%%%%%%%%%%%%%%%%%%%%%%%%%%%%%%%%%%%%%%
\section{Transversely polarized photons in vector-meson electroproduction}
%%%%%%%%%%%%%%%%%%%%%%%%%%%%%%%%%%%%%%%%%%%%%%%%%%%%%%%%%%%%%%%%%%%%%%%%%%%%%%%%%%%%%%%   
 
\begin{figure}[t]
\begin{center}
\includegraphics[width=0.48\tw,bb=103 462 388 658,clip=true]{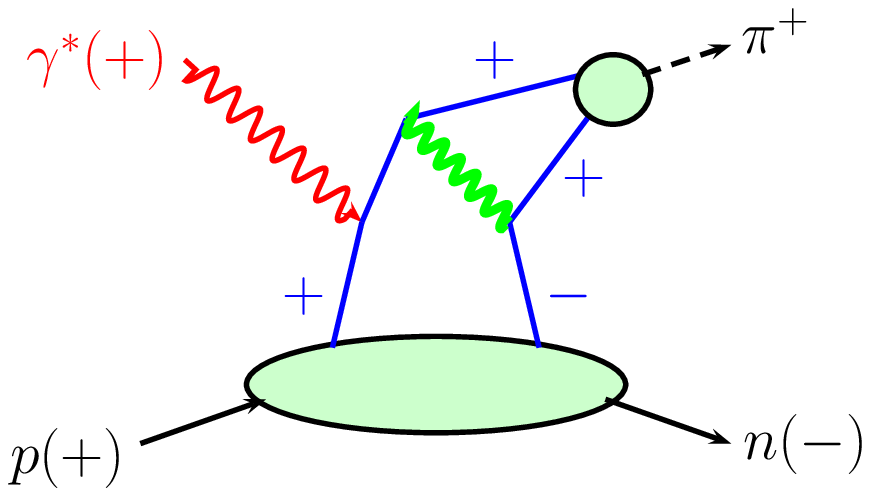}
\includegraphics[width=0.48\tw, bb= 39 350 526 727,clip=true]{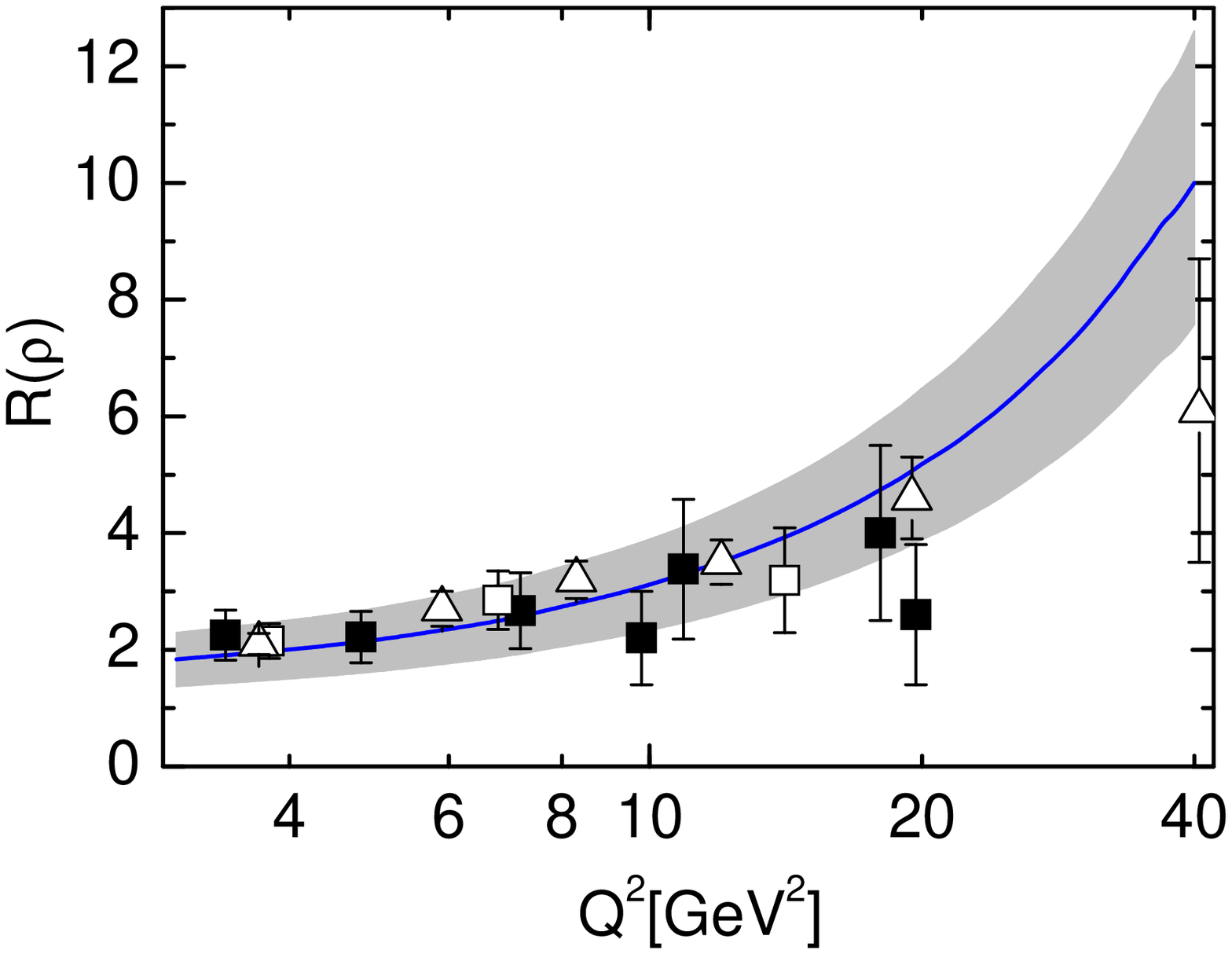}
\caption{\label{fig:0} A typical lowest order Feynman graph for meson electroproduction.
The signs indicate helicity labels for the contribution from transversity GPDs
to the amplitude ${\cal M}_{0-,++}$, see text.}
\caption{\label{fig:1} The ratio $R$ for $\rho^0$ production versus $Q^2$ at $W=90\,\gev$.
Data taken from H1~\protect\ci{H1-00} (filled symbols) and ZEUS 
\protect\ci{ZEUS07} (open symbols). The solid line represents the handbag results with 
the shaded bands indicating the theoretical uncertainties~\protect\ci{GK3}.}
\end{center}
\end{figure}

In a number of experiments, e.g.\ Refs.\ \ci{HERMES09,ZEUS07,H1-00}, the ratio of 
the longitudinal and transversal cross sections has been determined 
from the SDME $r_{00}^{04}$:
\be
R=\frac{\sigma_L}{\sigma_T} = \frac1{\varepsilon}\,\frac{r_{00}^{04}}{1-r_{00}^{04}}\,,
\label{eq:R}
\ee
where $\varepsilon$ is the ratio of the longitudinal and transversal photon fluxes. 
With regard to the factorization properties of meson electroproduction one
expects $R \propto Q^2$\,. In Fig.\ \ref{fig:1} the HERA data for $R$ are
displayed. One observes that $R$ is not at all large. At $Q^2\simeq 4\,\gev^2$
it is about 2, i.e.\ $\sigma_L\simeq 2\sigma_T$ only. For larger $Q^2$  the
ratio seems to increase slowly. Evidently, there are substantial contributions 
from $\gamma_T^*\to V_T$ transitions to vector-meson electroproduction. 

In a series of papers~\ci{GK1,GK2,GK3} a handbag approach has
been advocated for in which the subprocess amplitudes are calculated within 
the modified perturbative approach~\ci{sterman}, and the 
GPDs are constructed from reggeized double distributions~\ci{mul94,rad98}. 
In this approach the quark transverse momenta are retained in the subprocess 
and Sudakov suppressions are taken into account. The partons are still emitted
and re-absorbed by the proton collinearly. For the meson wave functions
Gaussians in the variable $k_\perp^2/(\tau(1-\tau))$ are assumed with
transverse size parameters fitted to experiment~\ci{jakob}. The variable $\tau$
denotes the fraction of the meson's momentum the quark entering the meson,
carries. It is to be emphasized that the $\gamma^*_T\to V_T$ transitions which
are infrared divergent in collinear approximation, are regularized by the 
quark transverse momenta in the modified perturbative approach. 

With this model the available data on cross sections and SDME for $\rho^0$ and
$\phi$ production have been fitted in the kinematical range $Q^2\gsim 3\,\gev^2$, 
$W\gsim 5\,\gev$ (i.e.\ for small values of skewness $\xi\simeq x_{Bj}/2\,\lsim\, 0.1$) 
and for the squared invariant momentum transfer $-t^\prime = -t+t_0\, 
\lsim\, 0.6\,\gev^2$ where $t_0$ is the value of $t$ for forward
scattering. Good agreement with experiment is found. As an example the 
results obtained in Ref.\ \ci{GK1} for $R(\rho^0)$ are displayed in Fig.\
\ref{fig:1} and compared to experiment. The data are well described within
that approach and the increase of $R$ according to Eq.\ \req{eq:R} is clearly 
visible. Results of similar quality have been obtained for $\phi$ production. 
The analysis carried through in \ci{GK1,GK2,GK3} fix the GPD $H$ 
for quarks and gluons. The other GPDs do practically not contribute to the 
cross sections and SDME at small skewness.

In experiment, e.g.\ \ci{HERMES09,H1-00}, there have also been
observed small but clearly non-zero contributions from $\gamma^*_T\to V_L$ 
transitions for instance in the SDME $r_{00}^5$. Such transitions, which 
violate $s$-channel helicity conservation, are not yet understood in the 
handbag approach. They are suppressed by $\propto \sqrt{-t}/Q$ as compared 
to the leading amplitude ${\cal M}_{0+,0+}$. For the even stronger suppressed 
$\gamma^*_L\to V_T$ and $\gamma^*_T\to V_T$ transitions there is no indication 
in experiment.

%%%%%%%%%%%%%%%%%%%%%%%%%%%%%%%%%%%%%%%%%%%%%%%%%%%%%%%%%%%%%%%%%%%%%%%%%%%%%%%%%%%%%%%%%%%%%
\section{Target asymmetries}
%%%%%%%%%%%%%%%%%%%%%%%%%%%%%%%%%%%%%%%%%%%%%%%%%%%%%%%%%%%%%%%%%%%%%%%%%%%%%%%%%%%%%%%%%%%%
The electroproduction cross sections measured with a transversely or
longitudinally polarized target consist of many terms, each can be projected
out by a $\sin{\varphi}$ or $\cos{\varphi}$ moment where $\varphi$ is a linear 
combination of $\phi$, the azimuthal angle between the lepton and the hadron 
plane and $\phi_s$, the orientation of the target spin vector~\ci{sapeta}. In
Tab.\ \ref{tab:1} the features of some of these moments are displayed. As the 
dominant interference terms reveal the target asymmetries provide detailed 
information on the $\gamma^* p\to VB$ amplitudes and therefore on the
underlying dynamics that generates them.    

\begin{table*}[t]
  \renewcommand{\arraystretch}{1.2}
\begin{center}
  \caption{Features of the asymmetries for transversally and longitudinally
    polarized targets. The angle $\theta_\gamma$ describes the rotation in the 
    lepton plane from the direction of the incoming lepton to the virtual
    photon one; it is very small.}
  \begin{tabular}{|c|| c | c | c | c |}
  \hline     
   observable  & dominant &   amplitudes  & low $t^\prime$ \\
	  & interf. term  &   &  behavior \\[0.2em]   
  \hline
  $A_{UT}^{\sin(\phi-\phi_s)}$ &  LL  & ${\rm Im}\big[{\cal M}^*_{0-,0+}
			 {\cal M}_{0+,0+}\big]$ & $\propto \sqrt{-t^\prime}$   \\[0.2em]
  $A_{UT}^{\sin(\phi_s)}$ & LT  &  ${\rm Im}\big[{\cal M}^*_{0-,++}{\cal M}_{0+,0+}\big]$  & const.  \\[0.2em]
  $A_{UT}^{\sin(2\phi-\phi_s)}$ & LT & ${\rm Im}\big[{\cal M}^*_{0\mp,-+}
			       {\cal M}_{0\pm,0+}\big]$ &  $\propto t^\prime$\\[0.2em]
  $A_{UT}^{\sin(\phi+\phi_s)}$ & TT &  ${\rm Im}\big[{\cal M}^*_{0-,++}
			  {\cal M}_{0+,++}\big]$ & $\propto \sqrt{-t^\prime}$    \\[0.2em]
  $A_{UT}^{\sin(2\phi+\phi_s)}$ & TT &  $\propto \sin{\theta_\gamma}$ &  $\propto t^\prime$\\[0.2em]
  $A_{UT}^{\sin(3\phi-\phi_s)}$ & TT & ${\rm Im}\big[{\cal M}^*_{0-,-+}
			       {\cal M}_{0+,-+}\big]$ & $\propto (-t^\prime)^{(3/2)}$  \\[0.2em]
  \hline
  $A_{UL}^{\sin(\phi)}$   & LT & ${\rm Im}\big[{\cal M}^*_{0-,++} {\cal M}_{0-,0+}\big]$ & 
  $\propto \sqrt{-t^\prime}$   \\[0.2em]
  \hline
  \end{tabular}
 \end{center}
  \label{tab:1}
 \renewcommand{\arraystretch}{1.0}   
  \end{table*} 

\begin{figure}[t]
  \begin{center}
  \includegraphics[width=0.44\tw,bb=25 333 532 743,clip=true]{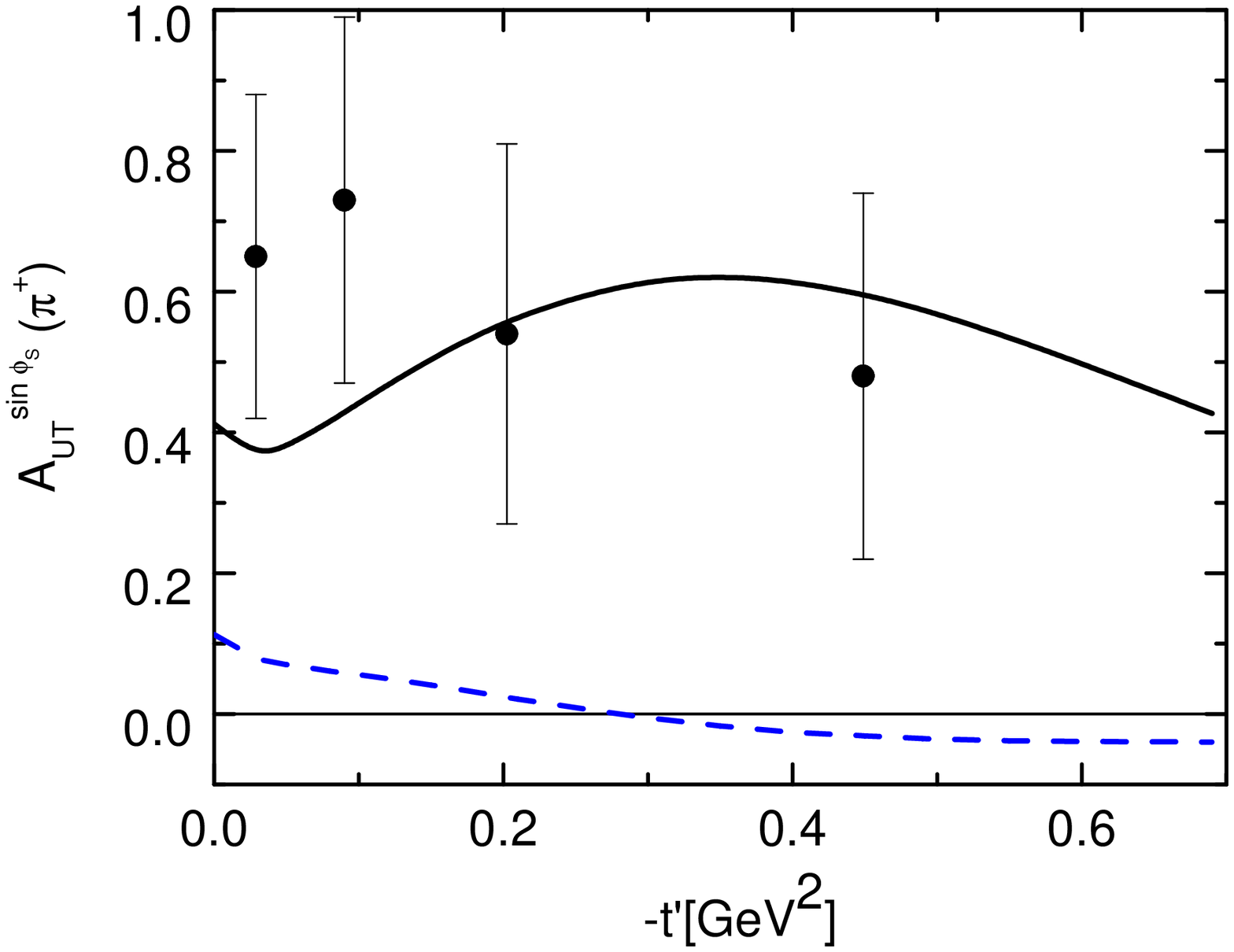}
\includegraphics[width=0.41\tw , bb= 44 327 533 753,clip=true]{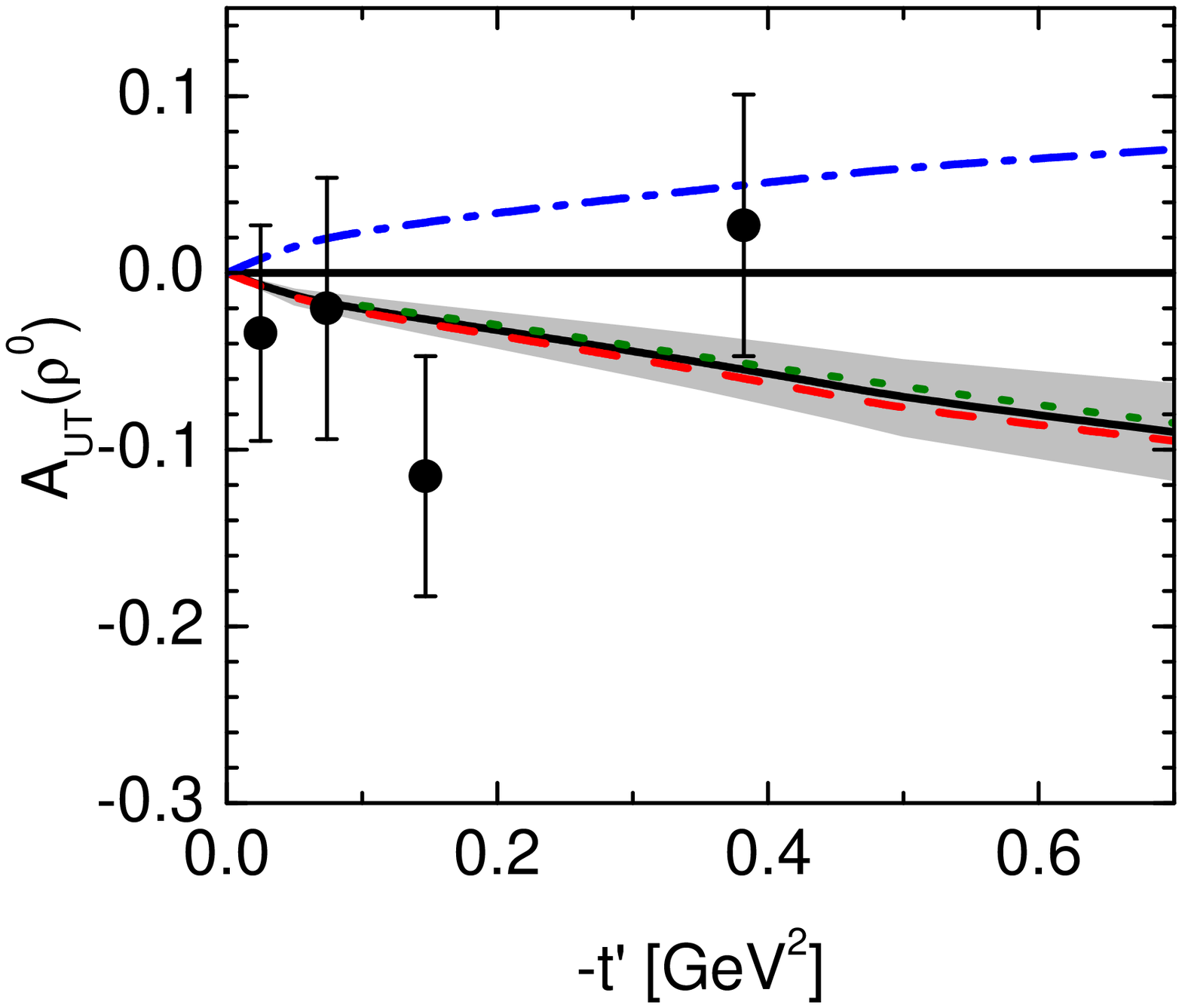}
  \caption{\label{fig:2} The $\sin{\phi_s}$ moment for a
    transversely polarized target at $Q^2\simeq 2.45\,\gev^2$ and
    $W=3.99\,\gev$ for $\pi^+$ production. The predictions from the handbag
    approach of Ref.\ \protect\ci{GK5} are shown as a solid line. The 
    dashed line is obtained disregarding the twist-3 contribution. Data are 
    taken from \protect\ci{Hristova}.}
\caption{\label{fig:3} The asymmetry $A_{UT}^{\sin{(\phi-\phi_s)}}$ for $\rho^0$ 
    production at $W=5\,\gev$ and $Q^2=2\,\gev^2$. Data taken from Ref.\ 
    \protect\ci{HERMES-rho}. The lines represent the results presented 
    in Ref.\ \protect\ci{GK4}. For further notations see text and Ref.\ 
    \protect\ci{GK4}.} 
\end{center}
\end{figure}

A number of these moments have been measured recently. A particularly striking
result is the $\sin{\phi_s}$ moment which has been measured by the HERMES 
collaboration for $\pi^+$ electroproduction~\ci{Hristova}. The data on this
moment, shown in Fig.\ \ref{fig:2}, exhibit a mild $t$-dependence and do not 
show any indication for a turnover towards zero for $t^\prime\to 0$. Inspection 
of Tab.\ \ref{tab:1} reveals that this behavior of $A_{UT}^{\sin{\phi_s}}$ at 
small $-t^\prime$ requires a contribution from the interference term 
${\rm Im}\big[{\cal M}_{0-,++}^*\,{\cal M}_{0+,0+}\big]$. Both the contributing 
amplitudes are helicity non-flip ones and are therefore not forced to vanish 
in the forward direction by angular momentum conservation. Thus, we see that 
also for pion electroproduction there are strong contributions  from 
$\gamma^*_T \to \pi$ transitions. The underlying dynamical mechanism for such
transitions will be discussed in Sect.\ 5.

For $\rho^0$ production the $\sin{(\phi-\phi_s)}$ moment has been measured by
HERMES~\ci{HERMES-rho} and COMPASS~\ci{COMPASS-rho}; 
the latter data being still preliminary. The HERMES data are shown in Fig.\ 
\ref{fig:3}. In the handbag approach $A_{UT}^{\sin{(\phi-\phi_s)}}$ can also
be expressed by an interference term between the convolutions of the GPDs $H$
and $E$ with hard scattering kernels 
\be
A_{UT}^{\sin{\phi-\phi_s}} \sim {\rm Im} \langle E \rangle^* \langle H \rangle
\ee
instead of the helicity amplitudes. Given that $H$ is known from the analysis
of the $\rho^0$ and $\phi$ cross sections and SDMEs, $A_{UT}$ provides 
information on $E$~\ci{GK3}. Let us recapitulate what we know about the GPD $E$.

%%%%%%%%%%%%%%%%%%%%%%%%%%%%%%%%%%%%%%%%%%%%%%%%%%%%%%%%%%%%%%%%%%%%%%%%%%%%%%%%%%%%%%%%%%%%
\section{The GPD $E$}
%%%%%%%%%%%%%%%%%%%%%%%%%%%%%%%%%%%%%%%%%%%%%%%%%%%%%%%%%%%%%%%%%%%%%%%%%%%%%%%%%%%%%%%%%%%
In Ref.\ \ci{DFJK4} the electromagnetic form factors of the proton and
neutron have been utilized in order to determine the zero-skewness GPDs for 
valence quarks through the sum rules which for the case of the Pauli form
factor, reads
\be
F_2^{p(n)} = \int_{0}^{1} dx \Big[e_{u(d)}\,E^u_v(x,\xi=0,t) + e_{d(u)}\,E^d_v(x,\xi=0,t)\Big]\,.
\ee 
In order to determine the GPDs from the integral a parameterization of the GPD
is required for which the ansatz  
\be
E_v^a(x,0,t)= e_v^a(x) \exp\Big[t(\alpha^\prime_v \ln(1/x)+b_e^a)\Big]
\label{E-ansatz}
\ee
is made in a small $-t$ approximation~\ci{DFJK4}.  The forward limit of $E$ is
parameterized analogously to that of the usual parton distributions:
\be
e_v^a = N_a x^{\alpha_v(0)} (1-x)^{\beta_v^a}\,,
\label{ev-ansatz}
\ee
where $\alpha_v(0)$ ($\simeq 0.48$) is the intercept of a standard Regge
trajectory and $\alpha_v^\prime$ in Eq.\ \req{E-ansatz} its slope. The 
normalization $N_a$ is fixed from the moment 
\be
\kappa^a = \int dx E^a_v(x,\xi,t=0)\,,
\ee
where $\kappa^a$ is the contribution of flavor-$a$ quarks to the anomalous
magnetic moments of the proton and neutron ($\kappa^u=1.67$, $\kappa^d=-2.03$). 
A best fit to the data on the nucleon form factors provides the powers 
$\beta_v^u=4$ and $\beta_v^d=5.6$. However, other powers are not excluded 
in the 2004 analysis of Ref.\ \ci{DFJK4}; the most extreme set of powers,
still in agreement with the form factor data, is $\beta_v^u=10$ and $\beta_v^d=5$. 
The analysis performed in \ci{DFJK4} should be repeated since 
new form factor data are available from Jefferson Lab, e.g.\ $G_{E}^n$ and 
$G_M^n$ are now measured up to $Q^2= 3.5$ and $5.0\,\gev^2$,
respectively~\ci{GEn,GMn}. These new data seem to favor $\beta_v^u<\beta_v^d$. 
The zero-skewness GPDs $E_v$ are used as input to a double distribution from 
which the valence quark GPDs for non-zero skewness are constructed~\ci{GK4}.

In Ref.\ \ci{GK4}, following Diehl and Kugler~\ci{kugler},
$E$ for gluons and sea quarks has been estimated from positivity bounds and a
sum rule for the second moments of $E$ which follows from a combination of
Ji's sum rule~\ci{ji97} and the momentum sum rule of deep
inelastic lepton-nucleon scattering. It has turned out that the valence quark 
contribution to that sum rule is very small, in particular if 
$\beta^u_v<\beta^d_v$, with the consequence of an almost exact cancellation 
of the gluon and sea quark moments. The GPDs $E^g$ and $E^{\rm sea}$ are 
parameterized analogously to $E_v$, see Eqs.\ \req{E-ansatz}, \req{ev-ansatz}. 
The normalization of $E^{\rm sea}$ is fixed by assuming that an appropriate 
positivity bound~\ci{poby,burkardt} is saturated while 
that of $E^g$ is determined from the sum rule. Several variants of $E$ have 
been exploited in Ref.\ \ci{GK4} in a calculation of $A_{UT}^{\sin{(\phi-\phi_s)}}$ 
within the handbag approach.  The results for a few variants are compared 
to the HERMES data on $\rho^0$ production~\ci{HERMES-rho} in 
Fig.\ \ref{fig:3}. Agreement between theory and experiment is to be noted. 
Similar agreement is obtained for the preliminary  COMPASS data~\ci{COMPASS-rho}. 
Combining both the experiments a negative value of 
$A_{UT}^{\sin{(\phi-\phi_s)}}$ for $\rho^0$ production is favored in agreement
with the theoretical results obtained in \ci{GK4}, only the 
extreme variant $\beta_v^u=10$ and $\beta_v^d=5$ (dashed-dotted line in Fig.\
\ref{fig:3}) seems to be ruled out. In Ref.\ \ci{GK4} predictions for 
$\omega$, $\rho^+$, $K^{*0}$ and $\phi$ productions are also given. Their 
comparison with forthcoming data from HERMES and COMPASS may provide
valuable restrictions on the GPD $E$.

With $E$ at hand one may exploit Ji's sum rule for the parton angular
momenta. At zero skewness the sum rule reads     
\be
\langle J^a\rangle = \frac12\big[ q^a_{20} + e^a_{20}\big]\,, \qquad
 \langle J^g\rangle = \frac12\big[ q^g_{20} + e^g_{20}\big]\,.
\ee
From a variant with $\beta_v^u=4$, $\beta_v^d=5.6$ and neglected $E^g$ 
and $E^{\rm sea}$ (solid line in Fig.\ \ref{fig:3}) for instance one obtains 
\be
\langle J^u \rangle = 0.250\,, \quad \langle J^d \rangle = 0.020\,, \quad 
\langle J^s \rangle = 0.015\,, \quad \langle J^g \rangle = 0.214\,, 
\ee
at the scale of $4\,\gev^2$. The angular momenta sum up to $\simeq 1/2$, the
spin of the proton. A very characteristic stable pattern is obtained in  
\ci{GK4}: For all variants investigated, $J^u$ and $J^g$ are large while 
the other two angular momenta are very small. The angular momenta of the 
valence quarks are $\langle J^u_v\rangle=0.222$ and $\langle
J^d_v\rangle=-0.015$. These values are identical to the results quoted in 
\ci{DFJK4} (for variant 1). They are also in agreement with a recent 
lattice result~\ci{lattice}.

%%%%%%%%%%%%%%%%%%%%%%%%%%%%%%%%%%%%%%%%%%%%%%%%%%%%%%%%%%%%%%%%%%%%%%%%%%%%%%%%%%%%%
\section{Target spin asymmetries in $\pi^+$ production}
%%%%%%%%%%%%%%%%%%%%%%%%%%%%%%%%%%%%%%%%%%%%%%%%%%%%%%%%%%%%%%%%%%%%%%%%%%%%%%%%%%%%%
In Ref.\ \ci{GK5} electroproduction of positively charged pions has been
investigated in the same handbag approach as applied to vector meson 
production~\ci{GK1,GK2,GK3} To the asymptotically leading amplitudes for 
longitudinally polarized photons the GPDs $\widetilde{H}$ and $\widetilde{E}$ 
contribute in the isovector combination 
\be
\widetilde{F}^{(3)}=\widetilde{F}^u_v-\widetilde{F}^d_v\,.
\ee
instead of $H$ and $E$ for vector mesons. In deviation to work performed in 
collinear approximation the full electromagnetic form factor of the pion as 
measured by the $F_\pi-2$ collaboration~\ci{horn06} is naturally 
taken into account~\footnote
{As compared to other work $\widetilde{E}$ contains only the non-pole contribution.}
(see also the recent work by Bechler and Mueller, Ref.\ \ci{bechler}). 
The GPDs $\widetilde{H}$ and $\widetilde{E}$ are again constructed with the 
help of double distributions with the forward limit of $\widetilde{H}$ being 
the polarized parton distributions while that of $\widetilde{E}$ is 
parameterized analogously to Eq.\ \req{ev-ansatz}
\be
\tilde{e}^u = - \tilde{e}^d = \widetilde{N}_e x^{-0.48} (1-x)^5\,.
\ee
The normalization $\widetilde{N}_e$ is fitted to experiment.

As is mentioned in Sect.\ 2 experiment requires a strong contribution from the
helicity-non-flip amplitude ${\cal M}_{0-,++}$ which does not vanish in the
forward direction. How can this amplitude be modeled in the frame work of the 
handbag approach? From the usual helicity non-flip GPDs $H, E, \ldots$ one
obtains a contribution to ${\cal M}_{0-,++}$ that vanishes $\propto t^\prime$
if it is non-zero at all. However, there is a second set of GPDs, the 
helicity-flip or transversity ones $H_T, E_T, \ldots$ \ci{diehl01,hoodbhoy}. As 
inspection of Fig.\ \ref{fig:0} where the helicity configuration of the
process is specified, reveals the proton-parton vertex is of non-flip nature
in this case and, hence, is not forced to vanish in the forward direction by 
angular momentum conservation. One also sees from Fig.\ \ref{fig:0}, that the 
helicity configuration of the subprocess is the same as for the full amplitude. 
Therefore, also the subprocess amplitude has not to vanish in the forward 
direction and so the full amplitude. The prize to pay is that quark and 
antiquark forming the pion have the same helicity. Therefore, the twist-3 pion 
wave function is needed instead of the familiar twist-2 one. The dynamical 
mechanism building up the amplitude ${\cal M}_{0-,++}$ is so of twist-3 order 
consisting of leading-twist helicity-flip GPDs and the twist-3 pion wave
function. This mechanism has been first proposed in \ci{passek} 
for photo- and electroproduction of mesons where $-t$ is considered as the 
large scale~\ci{huang}.

In Ref.\ \ci{GK5} the twist-3 pion wave function is taken from  
\ci{braun90} with the three-particle Fock component neglected. This wave 
function, still containing a pseudoscalar and a tensor component, is
proportional to the parameter $\mu_\pi=m^2_\pi/(m_u+m_d) \simeq 2\,\gev$ 
at the scale of $2\,\gev$ as a consequence of the divergency of the
axial-vector current ($m_u$ and $m_d$ are current quark masses). It is 
further assumed that the dominant transversity GPD is $H_T$ while the other
three can be neglected. The forward limit of $H^a_T$ is the transversity 
distribution $\delta^a(x)$ which has been determined in \ci{anselmino} 
in an analysis of data on the asymmetries in semi-inclusive electroproduction
of charged pions measured with a transversely polarized target. Using these
results for $\delta^a(x)$  the GPDs $H_T^a$ have been modeled in a manner 
analogous to that of the other GPDs ( see Eq.\ \req{E-ansatz})~\footnote
{While the relative signs of $\delta^u$ and $\delta^d$ is fixed in the
  analysis performed by Ref.\ \ci{anselmino} the absolute sign is not. 
  Here, in $\pi^+$ electroproduction a positive $\delta^u$ is required by 
  the signs of the target asymmetries.}.

\begin{figure}[t]
  \begin{center}
  \includegraphics[width=0.44\tw,bb=16 349 533 743,clip=true]{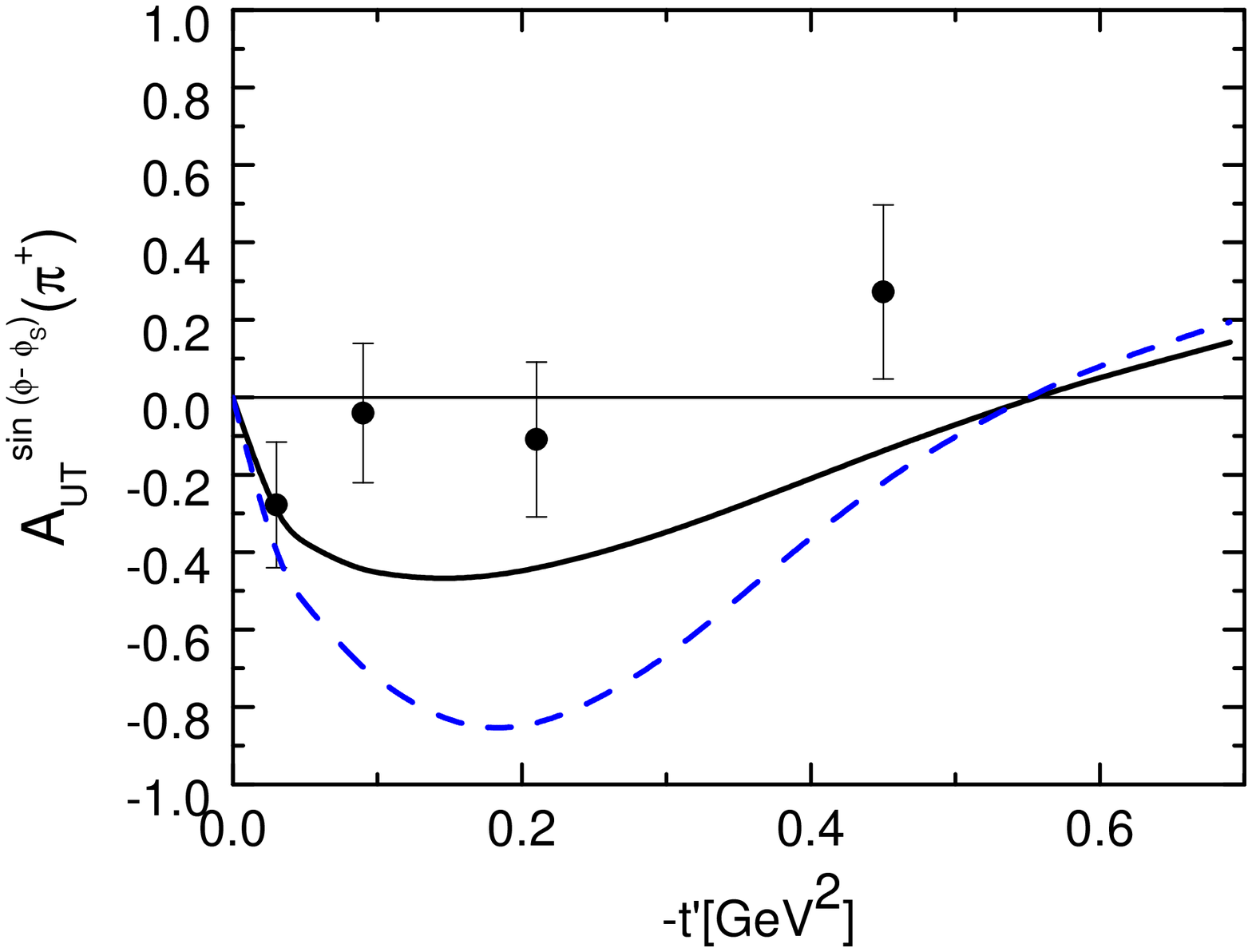}
  \includegraphics[width=0.42\tw,bb=30 346 533 746,clip=true]{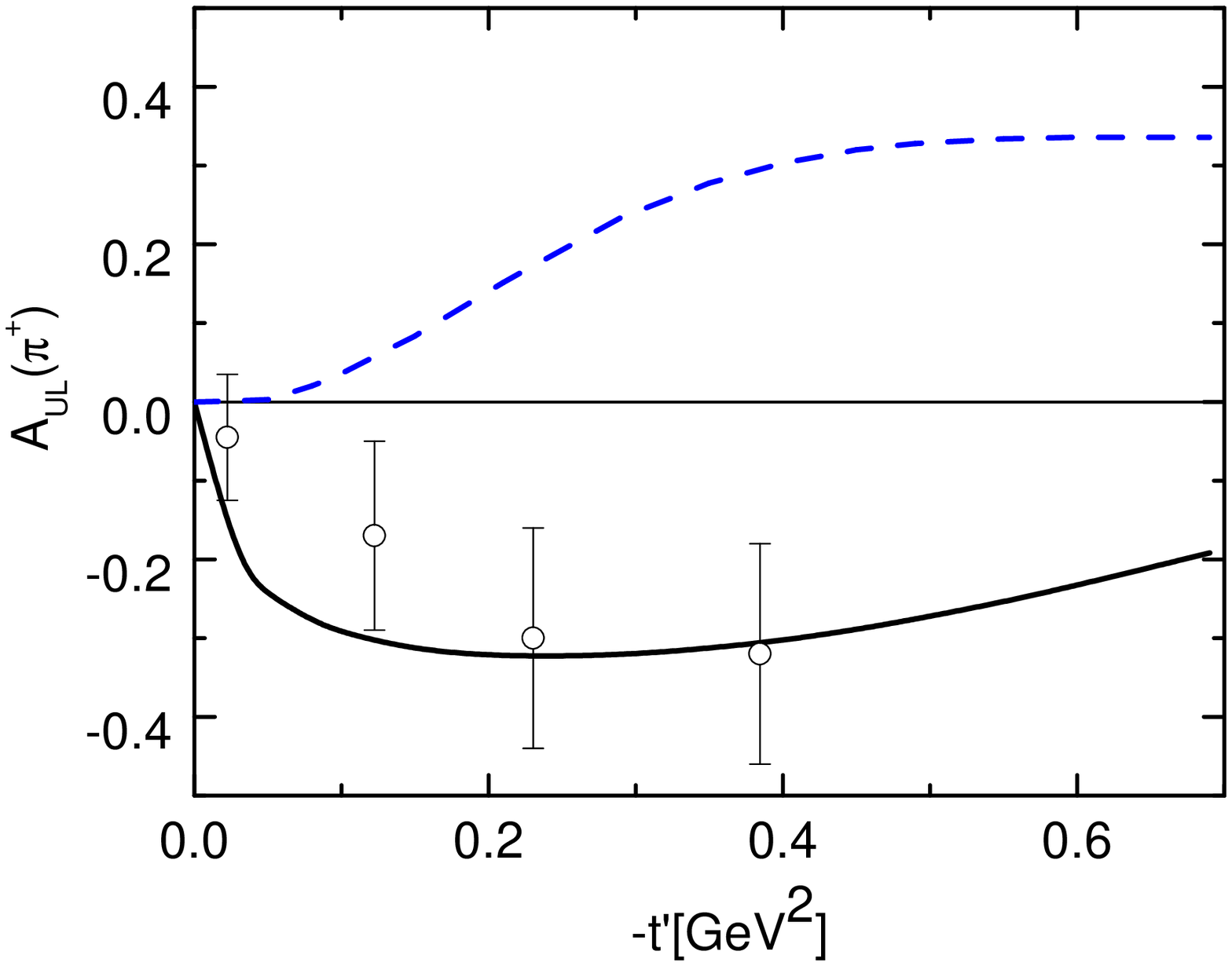}
  \caption{ \label{fig:5} Left: Predictions for the  $\sin{(\phi-\phi_s)}$ moment at 
  $Q^2=2.45\,\gev^2$  and $W=3.99\,\gev$ shown as solid lines~\protect\ci{GK5}. The dashed line 
   represents the longitudinal contribution to the $\sin{(\phi-\phi_s)}$ moment. Data are taken 
   from \protect\ci{Hristova}.}
  \caption{\label{fig:6} Right:  The asymmetry for a longitudinally polarized target at
  $Q^2\simeq 2.4\,\gev^2$ and $W\simeq 4.1\,\gev$. The dashed line is obtained disregarding the 
  twist-3 contribution. Data are taken from \protect\ci{hermes02}.}
  \end{center}
  \end{figure}

It is shown in \ci{GK5} that with the described model GPDs, the 
$\pi^+$ cross sections as measured by HERMES~\ci{HERMES07} are 
nicely fitted  as well as the transverse target asymmetries~\ci{Hristova}. This 
can be seen for $A_{UT}^{\sin{\phi_s}}$ from Fig.\ \ref{fig:1}. Also the 
$\sin(\phi-\phi_s)$ moment which is dominantly fed by an interference term 
of the the two amplitudes for longitudinally polarized photons (see Tab.\ 
\ref{tab:1}), is fairly well described, as is obvious from Fig.\
\ref{fig:5}. Very interesting is also the asymmetry for a longitudinally 
polarized target which is dominated by the interference term between 
${\cal M}_{0-,++}$ which comprises the twist-3 effect, and the nucleon 
helicity-flip amplitude for $\gamma^*_L\to \pi$ transition, ${\cal M}_{0-,0+}$. 
Results for $A_{UL}^{\sin \phi}$ are displayed in Fig.\ \ref{fig:6} and compared 
to the data~\ci{hermes02}. Also in this case good agreement between theory 
and experiment is to be seen. In both the cases, $A_{UT}^{\sin{\phi_s}}$ 
and $A_{UL}^{\sin \phi}$ the prominent role of the twist-3 mechanism is
clearly visible. Switching it off one obtains the dashed lines which are
significantly at variance with experiment. In this case the transverse 
amplitudes are only fed by the pion-pole contribution. The other transverse
target asymmetries quoted in Tab.\ \ref{tab:1} are predicted to be small 
in absolute value which is in agreement with experiment~\ci{Hristova}.
Thus, in summary, there is strong evidence for transversity in hard exclusive
pion electroproduction. It should be considered as a non-trivial result that 
the transversity distributions determined from data on inclusive processes 
lead to a transversity GPD which is nicely in agreement with target 
asymmetries measured in exclusive pion electroproduction. 

It is to be stressed that information on the amplitude ${\cal M}_{0-,++}$ can
also obtained from the asymmetries measured  with a longitudinally polarized
beam or with a longitudinally polarized beam and target. The first asymmetry, 
$A_{LU}^{\sin \phi}$, is dominated by the same interference term as $
A_{UL}^{\sin\phi}$ but diluted by the factor $\sqrt{(1-\varepsilon)/(1+\varepsilon)}$. 
Also the second asymmetry, $A_{LL}^{\cos \phi}$, is dominated 
by the interference term ${\cal M}_{0-,++}^*\,{\cal M}_{0-,0+}$. However, in
this case its real part occurs. For HERMES kinematics it is predicted to be
rather large and positive at small $-t^\prime$ and changes sign at 
$-t^\prime \simeq 0.4\,\gev^2$ \ci{GK5}. A measurement of these asymmetries
would constitute a serious check of the twist-3 effect.  
 
Although the main purpose of the work presented in \ci{GK5} is
focused on the analysis of the HERMES data one may be also interested in
comparing this approach with the Jefferson Lab data on the cross sections~\ci{horn06}. 
With the GPDs $\widetilde{H}, \widetilde{E}$ and $H_T$ in their present form
the agreement with these data is reasonable for the transverse cross section 
while the longitudinal one is somewhat too small. It is however to be stressed 
that the approach advocated for in \ci{GK1,GK5,GK4} is designed
for small skewness. At larger values of it the parameterizations of the GPDs 
are perhaps to simple and may require improvements. It is also important to 
realize that the GPDs are probed by the HERMES, COMPASS and HERA data only at 
$x$ less than about 0.6. One may therefore change to some extent the GPDs 
for large $x$ without changing the results for cross sections and asymmetries
in the kinematical region of small skewness. For Jefferson Lab kinematics, on 
the other hand, such changes of the GPDs may matter. 

  %%%%%%%%%%%%%%%%%%%%%%%%%%%%%%%%%%%%%%%%%%%%%%%%%%%%%%%%%%%%%%%%%%%%%%%%%%%%
  \section{Summary}
  \label{sec:summary}
  %%%%%%%%%%%%%%%%%%%%%%%%%%%%%%%%%%%%%%%%%%%%%%%%%%%%%%%%%%%%%%%%%%%%%%%%%%%
Recent measurements of spin effects in hard meson electroproduction has been
reviewed. The spin effects include separated electroproduction cross 
sections, SDME and target as well as beam asymmetries. The data clearly show 
that a leading-twist calculation of meson electroproduction within the handbag 
approach is insufficient. They demand higher-twist and/or power corrections 
which manifest themselves through substantial contributions from $\gamma_T^*$ 
to meson transitions.

A most striking effect is the target asymmetry $A_{UT}^{\sin \phi_s}$ in
$\pi^+$ electroproduction. The interpretation of this effect requires a large
contribution from the helicity non-flip amplitude ${\cal M}_{0-,++}$. Within 
the handbag approach such a contribution is generated by the helicity-flip or 
transversity GPDs in combination with a twist-3 pion wave function~\ci{GK5}.
This explanation establishes an interesting connection to transversity
parton distributions measured in inclusive processes. Further studies of 
transversity in exclusive reactions are certainly demanded. For instance, 
data on the asymmetries obtained with a longitudinally polarized beam and 
with likewise polarized beam and target would be very helpful in settling 
this dynamical issue.

Good data on $\pi^0$ electroproduction would also be highly welcome. They
would not only allow for an additional test of the twist-3 mechanism but also
give the opportunity to verify the model GPDs $\widetilde{H}$ and 
$\widetilde{E}$ as used in \ci{GK5}.% and therefore implicitly to 
%probe the parameterization of the pion-pole contribution.
  
One may wonder whether the twist-3 mechanism does not apply to vector-meson 
electroproduction as well and offers an explanation of the experimentally 
observed $\gamma_T^*\to V_L$ transitions mentioned in Sect.\ 2.  
It however turned out that this effect is too small in comparison to the data, for
instance, $r_{00}^{05}$. The reason is that instead of the parameter 
$\mu_\pi$ the mass of the vector meson sets the scale of the twist-3 effect. 
This amounts to a reduction by about a factor of three. Further suppression 
comes from the unfavorable flavor combination of $H_T$ occurring for 
uncharged vector mesons, e.g.\ $e_u H_T^u-e_d H_T^d$ for $\rho^0$ production 
instead of $H_T^u-H_T^d$ for $\pi^+$ production. Perhaps the gluonic GPD 
$H_T^g$ may lead to a larger effect.
  
  %%%%%%%%%%%%%%%%%%%%%%%%%%%%%%%%%%%%%%%%%%%%%%%%%%%%%%%%%%%%%%%%%%%%%%%%%%%%
  {\bf Acknowledgements} This work is supported  in part by the Heisenberg-Landau
  program and by the BMBF, contract number 06RY258.

  \vskip 10mm 
  %%%%%%%%%%%%%%%%%%%%%%%%%%%%%%%%%%%%%%%%%%%%%%%%%%%%%%%%%%%%%%%%%%%%%%
  
  \end{document}